\documentstyle[prl,aps,multicol,epsf,rotate]{revtex}

\begin{document}
\draft

\title{Theory of Branching and Annihilating Random Walks}

\author{John Cardy and Uwe C. T\"auber}

\address{University of Oxford, Department of Physics -- Theoretical
	Physics, 1 Keble Road, Oxford OX1 3NP, U.K.}

\date{\today}

\maketitle

\begin{abstract}
A systematic theory for the diffusion--limited reaction processes 
$A + A \to \oslash$ and $A \to (m+1) A$ is developed. Fluctuations are
taken into account via the field--theoretic dynamical renormalization
group. For $m$ even the mean field rate equation, which predicts only
an active phase, remains qualitatively correct near $d_c = 2$
dimensions; but below $d_c' \approx 4/3$ a nontrivial transition to an
inactive phase governed by power law behavior appears. For $m$ odd
there is a dynamic phase transition for any $d \leq 2$ which is
described by the directed percolation universality class.
\end{abstract}


\begin{multicols}{2}

Nonequilibrium models with an extensive number of degrees of freedom
whose dynamics violates detailed balance occur in studies of many
biological, chemical and physical systems. Like equilibrium systems,
their stationary states may exhibit phase transitions which in many
cases appear to fall into distinct classes characterized by universal
quantities such as critical exponents. One of the most common such
classes is that exemplified by {\it directed percolation} (DP)
\cite{dirper}. This represents a transition from a nontrivial `active'
steady state to an absorbing `inactive' state with no fluctuations. 
Many nonequilibrium phase transitions appear to belong to this
universality class, e.g., the contact process \cite{contct}, the dimer
poisoning problem in the ZGB model \cite{zgbmod}, and auto--catalytic
reaction models \cite{autcat}. The universal properties of the DP
transition are theoretically well understood in the context of a
renormalization group (RG) analysis based on an expansion around mean
field theory below the upper critical dimension $d_c = 4$
\cite{dpftrg}.

More recently a class of models has been studied which, in certain
cases, appear as exceptions to the general rule that such transitions
should fall into the DP universality class. These include a
probabilistic cellular automaton model \cite{celaut}, certain kinetic
Ising models \cite{kising,norsim}, and an interacting monomer--dimer
model \cite{mondim}. In one dimension the dynamics of these is
equivalent to a class of models called {\it branching and annihilating
random walks} (BARWs) \cite{taktre,jensen,redner}, which also have a
natural generalization to higher dimensions. In the language of
reaction--diffusion systems, BARWs describe the stochastic dynamics of
a single species of particles $A$ undergoing three basic processes:  
diffusion, often modeled by a random walk on a lattice and
characterized by a diffusion coefficient $D$; an annihilation reaction
$A + A \to \oslash$ when particles are close (or on the same site), at
rate $\lambda$; and a branching process $A \to (m+1) A$ (where $m$ is
a positive integer), at rate $\sigma_m$. The above--mentioned
one--dimensional models all correspond to the case $m = 2$. For the 
kinetic Ising model, the particles $A$ are to be identified with the
domain walls, and the transition to the inactive state corresponds to
the ordering of the Ising spins \cite{kising,norsim}. In general, this
new universality class has been observed in $d = 1$ for {\it even}
values of $m$, when the number of particles is locally conserved
modulo 2. When $m$ is {\it odd}, the DP values of the exponents appear
to be realized. (It should be remarked that several of the models
which have been studied do not contain three independent parameters
corresponding to $D$, $\lambda$, and $\sigma_m$ so that it may occur
that the actual transition is inaccessible. This appears to be so for
the simplest lattice BARW model with $m = 2$, which is always in the
inactive phase \cite{taktre}.)

Besides the appearance of a new universality class, another issue
which clearly requires theoretical explanation is the {\it occurrence}
of a transition at a finite value of $\sigma_m$. For the mean field
rate equation for the average density
\begin{equation}
 \label{mfrate}
	{\dot n}(t) = - 2 \lambda \, n(t)^2 + m \sigma_m \, n(t) 
\end{equation}
predicts a non-zero steady state density $m \sigma_m / 2 \lambda$, so
that this state should be 
{\it active} for all $\sigma_m > 0$, in contrast to what is in fact
observed for $d = 1$. It is, however, known that fluctuation effects
render the mean field description of the pure annihilation problem
($\sigma_m = 0$) qualitatively incorrect for $d \leq 2$ \cite{annlee}. 
Therefore, a detailed theory has to demonstrate how these are also
responsible for moving the critical value of $\sigma_m$ away from
zero.

In this Letter, we describe the first systematic theory of these 
phenomena (details will be presented elsewhere \cite{cartau}). It is
based on the field--theoretic RG analysis which has proven successful
for the DP problem \cite{dpftrg}: however, we correct an important
error which was made in an earlier investigation along the same lines
\cite{celaut}. A summary of our main results follows.

(a) {\it Even} $m$. For $d > 2$ the mean field description
(\ref{mfrate}) is qualitatively correct in that the transition occurs
at $\sigma_m = 0$. The density in the active phase vanishes as 
$n \propto \sigma_m$, with calculable logarithmic corrections in two
dimensions. As $d$ is lowered below 2, the transition first continues
to occur at $\sigma_m = 0$, with modified critical exponents, until a
{\it second critical dimension} $d_c' \approx 4/3$ is reached. Below
this, and in particular for $d = 1$, there appears a nontrivial
transition at $\sigma_c > 0$ from the active phase to an inactive
phase in which the density decays asymptotically as 
$n \propto t^{-d/2}$. Because of the existence of {\it two} critical
dimensionalities, this new universality class apparently has no simple
mean field limit, close to which the fluctuations can be controlled. 
We are therefore unable to generate a systematic $\epsilon$ expansion
for the critical exponents. The truncated loop expansion in fixed
dimension seems to provide at least a qualitative description of the
transition; however, as there exists no small expansion parameter, the
actual values for the critical exponents to one--loop order are rather
inaccurate. The RG analysis also shows that higher values of $m$
inevitably generate an effective $m = 2$ reaction under
renormalization, and this is always the most relevant term. Therefore
all such processes with even $m$ fall into the same universality
class. We have also considered an $N$ species generalization of the
$m = 2$ model. The RG analysis shows that for $N > 1$ the critical
behavior is that of the $N \to \infty$ limit, which is exactly
solvable but always in the {\it active} phase for $\sigma_2 > 0$, with
critical exponents being described by yet another universality class.

(b) {\it Odd} $m$. The case $m = 1$ here is typical. Under
renormalization, a spontaneous decay process $A \to \oslash$ is
generated for arbitrarily small branching rates, by the combined
reactions $A \to 2A$, $2A \to \oslash$, so that the density decays 
{\it exponentially}, as in the inactive phase of DP. We find this to
occur for $d < 2$, when the pure annihilation process is relevant. For
$d \geq 2$, however, the situation is more subtle. The RG analysis in
this case predicts that there is in fact a nontrivial transition even
at $d_c = 2$, with $\sigma_c \sim D e^{- 4 \pi D / \lambda}$, while it
is absent for $d > 2$. Analysis of the effective theory for $d \leq 2$
then shows that the subsequent transition at larger values of
$\sigma_1$ is in the DP universality class, as is observed in
simulations \cite{taktre,jensen}. 

The field--theoretic analysis of these problems begins from the
`second--quantized' approach to classical stochastic particle systems
which is well known and has been described in detail elsewhere
\cite{annlee,recdif}. Annihilation and creation operators $a_i$ and
$a_i^\dagger$, satisfying the usual boson commutation relations, are
introduced at each site $i$ of the lattice, and the time--dependent
state vector $| \Psi(t) \rangle \equiv \sum_{\{n_i\}} p(\{n_i\};t) 
\prod_i {a_i^\dagger}^{n_i} | 0 \rangle$ is constructed from the
probabilities $p(\{n_i\};t)$. The (classical) master equation
satisfied by these may then be recast as a Schr\"odinger--like 
equation with a time evolution operator which, in this example, has
the form $H = H_d + H_a + H_b + H_T$, where 
\begin{eqnarray}
 \label{hamilt}
	H_d &=& D \sum_{(ij)} \left( a_i^\dagger - a_j^\dagger \right)
				\left( a_i - a_j \right) 	\ , \\
	H_a &=& - \lambda \sum_i 
		\left( a_i^2 - {a_i^\dagger}^2 a_i^2 \right) 	\ , \\
	H_b &=& - \sigma_m \sum_i \left( {a_i^\dagger}^{m+1} a_i 
					- a_i^\dagger a_i \right)\ ,\\ 
	H_T &=& - \tau \sum_i \left( {a_i^\dagger}^2 - 1 \right) \ .
\end{eqnarray}
The last term corresponds to a constant creation of pairs of
particles, simulating the effects of finite temperature in the kinetic
Ising model \cite{norsim}. Finally, utilizing the coherent--state
path integral formalism, the `quantum many particle' hamiltonian
$H$ can be cast into a field theory which describes BARW processes
{\it including} fluctuation effects. Note that no additional
assumptions, specifically regarding the form of the noise correlations
in an extension of Eq.~(\ref{mfrate}) to an effective Langevin
equation, had to be invoked in this derivation.

When $m$ is {\it even}, there is a formal symmetry of $H$ under
changing the signs of all the $a_i$ and $a_i^\dagger$ simultaneously:
this corresponds to the conservation of particle number modulo 2. 
However, in this formalism, expectation values of operators such as
the local density $n_j = a_j^\dagger a_j$, for example, are given by
matrix elements of the form 
$\langle 0 | e^{\sum_i a_i} n_j e^{-Ht} | \Psi(0) \rangle$, and in
order to use time--dependent perturbation theory and Wick's theorem it 
is conventional to commute the factor $e^{\sum_i a_i}$ through. This
is equivalent to applying the formal shift 
$a_i^\dagger \to 1+ a_i^\dagger$ in $H$. Yet this obscures the above
symmetry, and if, in accordance with the usual naive power counting
arguments near the upper critical dimension, higher order quartic
terms in $H$ are then ignored, it becomes completely lost. This led
the authors of Ref.~\cite{celaut} to the erroneous conclusion that,
near $d = 4$, the transition should be in the DP universality class
{\it irrespective} of the parity of $m$.

However, it is imperative in any RG analysis to preserve all known
symmetries of the system. In the present case, this may be done by
observing that the RG equations themselves (as opposed to the
calculations of observables such as the density) should be independent 
of which basis is used, and it is therefore possible, and, indeed,
necessary, to perform the computations in the representation of the
model in which the symmetry is manifest. The methods for doing this
are standard, and will be described in detail elsewhere \cite{cartau}. 
The case $\sigma_m = \tau = 0$, corresponding to a pure annihilation
process, has already been analyzed in \cite{annlee}. The RG equation
for the flow of the dimensionless coupling 
$\ell \equiv C_d \lambda / D \kappa^\epsilon$, where $\kappa$ is a
normalization wave number, $C_d = \Gamma(2-d/2) / 2^{d-1} \pi^{d/2}$ a 
geometric factor, and $\epsilon = 2 - d$, under a rescaling factor
$e^l$, is given by $d \ell / d l = \epsilon \ell - \ell^2$, which is
exact at one loop. For $d < 2$ the late time behavior is controlled by
the nontrivial fixed point at $\ell^* = \epsilon$, leading to an 
asymptotic particle density decay according to 
$n(t) \propto t^{-d/2}$.

The first question to be addressed is whether the branching rate
$\sigma_m$ is relevant at the pure annihilation fixed point, i.e.,
whether its RG eigenvalue $y_\sigma$ is positive. If so, the late time
behavior must differ from that of the pure annihilation process,
indicating that the active phase is reached immediately. For 
$d \geq 2$ we find $y_\sigma = 2$ from simple power counting, so
indeed $\sigma_m$ is relevant. The density in the active phase
vanishes according to the mean field result $n \propto \sigma_m$. For
$d < 2$, to one--loop order, 
$y_\sigma = 2 - [m(m+1)/2] \ell + O(\ell^2)$, so that $\sigma_m$
remains relevant at the annihilation fixed point $\ell^* = \epsilon$
just below $d = 2$, with the lower values of $m$ the being most
relevant. Since these lower allowed values of $m$ inevitably become
generated whenever the annihilation rate is nonzero, we conclude that
the cases with $m$ even will always fall into the universality class
of $m = 2$, while $m$ odd will generate $m = 1$ and $m = -1$. The
latter is always relevant, and, as we shall see below, is responsible
for the crossover to the DP universality class. For the time being we
therefore restrict our attention to the case of even $m$. For $d = 2$
the marginality of $\ell$ is responsible for logarithmic corrections
to mean field theory, which for $m = 2$ take the form 
$n \propto \sigma / [\ln (1/\sigma)]^2$. 

The above result for $y_\sigma$ is valid only close to $d = 2$. 
Fortunately it is possible to compute it {\it exactly} in one
dimension, at the pure annihilation fixed point. The latter
corresponds to the limit of infinite bare coupling $\lambda$
\cite{annlee}. The multiparticle states then effectively propagate as
hard--core bosons in between the annihilation and branching processes,
and so, in one dimension, behave like free fermions. On the lattice,
this limit only makes sense if we define the branching process as
placing the $m$ offspring on different but neighboring sites. The
branching contribution to $H$, in terms of these fermionic operators
$c_i$ and $c_i^\dagger$, thus acquires the form 
$H_b = \sigma_m \sum_i \prod_{j=-m/2}^{m/2} c_{i+j}^\dagger c_i$. 
The continuum limit of this expression, found by performing a Taylor
expansion in powers of the lattice spacing $a_0$, will be different
from the bosonic case because the anticommuting nature of the
$c_i^\dagger$ allows each derivative to appear only {\it once}. The
lowest order term has the form 
$a_0^{m(m+1)/2} c^\dagger (\partial c^\dagger) (\partial^2 c^\dagger)
\ldots (\partial^m c^\dagger) c$, with the result that the effective
expansion parameter $\tilde \sigma_m \equiv a_0^{m(m+1)/2} \sigma_m$
has a modified scaling dimension. This leads to the result (which may
be confirmed by other less formal methods) that 
$y_\sigma = 2 - m(m+1)/2$ {\it exactly} in $d = 1$. Thus, for reasons
we do not understand, the $O(\epsilon)$ result appears to be exact in
$d=1$, and $y_\sigma$ changes sign at a value of 
$d = d_c' \approx 4/3$ for $m = 2$, if the higher order terms continue
to be small. In $d = 1$, $y_\sigma < 0$ for all the even values of
$m$. This establishes the result that the late time behavior for small
values of $\sigma_m$ is controlled by the annihilation fixed point, so
that $n(t) \propto t^{-1/2}$. In the inactive phase, the system is
composed of a set of highly {\it anticorrelated} bunches of odd
numbers of particles, the spatial distribution of which, upon coarse
graining, looks like that of single particles in the pure annihilation
process.
 
Clearly, the above scenario cannot be obtained in any finite order of
an expansion near $d_c = 2$. We have therefore performed a truncated
loop expansion at fixed dimension, retaining the full dependence on
$\sigma$, which appears both as a vertex and as a mass term. To one
loop order, the RG flow equations for the renormalized reaction rates
$\ell = C_d \lambda / D \kappa^{2-d}$ and $s = \sigma / D \kappa^2$
read ($m = 2$)
\begin{eqnarray}
 \label{annihi}
	d \ell / d l &=& \ell [ 2 - d - \ell / (1 + s)^{2-d/2} ] \ , \\
 \label{branch}
	d s / d l &=& s [ 2 - 3 \ell / (1 + s)^{2-d/2} ] \ . 
\end{eqnarray}
For $s \to 0$, the annihilation fixed point $\ell^* = 2 - d$ of the
inactive phase is recovered, while for $s \to \infty$ the loop 
contributions to the anomalous dimensions vanish, and the flow
approaches a Gaussian fixed point describing the active phase. For
large $s$, the effective coupling in Eqs.~(\ref{annihi}),
(\ref{branch}) becomes $g \equiv \ell / s^{2-d/2}$ 
(see Sec.~III of Ref.~\cite{tausch}), whose flow is given by 
$d g / d l = 2 g - [(10-3d)/2] g^2 \equiv - \beta(g)$. In addition to
the stable Gaussian fixed point at $g = 0$ there is a nontrivial
{\it unstable} one at $g^* = 4 / (10 - 3d)$ describing the phase
transition. At this order there is neither field nor diffusion constant
renormalization, giving a dynamic exponent $z \approx 2$. However,
because the mean field density $n \sim \sigma / \lambda$ and the
spatial correlation length $\xi_\perp\sim\sigma^{-1/2}$ depend not
just on $g$ but also on the dangerous irrelevant variable $s^{-1}$,
the critical exponents describing the approach to the critical point
in the active phase depend not only on 
$y_\varepsilon \equiv \beta'(g^*)$, describing the distance from the
critical point $\varepsilon \equiv (g^* - g) / g^*$, but also on 
$y_\lambda \approx 2 - d - g^*$ and $y_\sigma \approx 2 - 3 g^*$. As a
result we find that $n \sim \varepsilon^\beta$, with 
$\beta = (d+y_\lambda-y_\sigma) / y_\varepsilon \approx 4 / (10-3d)$, 
and $\xi_\perp \sim \varepsilon^{-\nu_\perp}$, with 
$\nu_\perp = (1-y_\sigma) / y_\varepsilon \approx 3 / (10-3d)$. The
truncated one--loop approximation thus seems to provide a
qualitatively correct picture of the transition, although the actual 
numerical values of these exponents in one dimension are rather poor
as compared to simulation results \cite{norsim}; this is not very
surprising, however, as there is no small expansion parameter present
here. In addition, we cannot really access those exponents that
describe the behavior {\it at} the critical point, as the density
might depend nonanalytically on $\sigma$ there; this also precludes a 
sound derivation of scaling relations \cite{norsim} connecting these
with the above exponents describing the active phase.

A better result is obtained for the exponent $\nu_\tau \equiv 1/y_\tau$
describing the divergence of the correlation length as the pair
creation rate $\tau \to 0$ at the critical point $\varepsilon = 0$. The
one--loop flow equation for $\tau$ reads
\begin{equation}
 \label{paircr}
	d\tau / dl = \tau [d+2 - \ell / (1+s)^{2-d/2}] \ ,
\end{equation}
and hence in the inactive phase, or at the critical point $\sigma = 0$ 
for $d > d_c'$, one has $\nu_\tau = 1 / 2d$, while at the nontrivial
phase transition for $d < d_c'$ the result is 
$\nu_\tau \approx (10-3d) / (16+4d-3d^2)$. In one dimension, 
$\nu_\tau \approx 7/17$, which is in fair agreement with simulations
\cite{norsim}.

We have also investigated an $N$ species generalization of the $m = 2$
problem, defined by the processes $2 A^\alpha \to \oslash$, at rate 
$\lambda / N$, $A^\alpha \to 3 A^\alpha$, at rate $\sigma$, and 
$A^\alpha \to A^\alpha + 2 A^\beta$, $\beta \not= \alpha$, at rate
$\sigma' / (N-1)$. To one--loop order at the annihilation fixed point
the RG eigenvalue of the additional branching process becomes 
$y_{\sigma'} = 2 - \ell$, which is therefore {\it more relevant} than
the original reaction with rate $\sigma$. We have chosen the above $N$
component version, because for $N \to \infty$, the ensuing theory 
(with $\sigma = 0$) can be solved exactly; physically this limit
corresponds to the situation where each particle may annihilate only
with its sibling. The resulting critical point remains at 
$\sigma_c = 0$ for all $d$, and its universality class, {\it distinct}
from the previously discussed ones, is characterized by the
mean--field exponents $z = 2$, $\beta = 1$, and as a consequence of
the now {\it exact} result $y_{\sigma'} = 2 - \ell$, we find 
$\nu_\perp \equiv 1 / y_{{\sigma'}} = 1/d$, using $\ell^* = 2 - d$ for 
$d \leq 2$. 

We now return to the case of odd $m$. As argued above, fluctuations
generate a spontaneous single--particle decay process, and the 
{\it effective} interactions at a given site acquire the form
\begin{equation}
 \label{heffns}
	\mu \left( a^\dagger - 1 \right) a
	- \sigma \left( {a^\dagger} - 1 \right) a^\dagger a
	+ \lambda \left( {a^\dagger}^2 - 1 \right) a^2 \, .
\end{equation}
When the single--particle decay rate $\mu \not= 0$, it is convenient
to remove the linear term in $a$ by the shift 
$a^\dagger \to 1 + a^\dagger$ mentioned earlier. This results in the
interaction hamiltonian
\begin{equation}
 \label{heffsh}
	{\tilde H}_{\rm eff}^{\rm int}=(\mu-\sigma) \, a^\dagger a 
	- \sigma \, {a^\dagger}^2 a + 2 \lambda \, a^\dagger a^2 
	+ \lambda \, {a^\dagger}^2 a^2 \ .
\end{equation}
If we now neglect the quartic term (justifiable in this case since
there is no `parity' symmetry that must be preserved), we find
precisely the interaction hamiltonian used to characterize DP
\cite{dpftrg}. The transition occurs when the renormalized version of
the mass term $\mu_R - \sigma_R$ vanishes. The question of whether
this actually happens for allowed values of the bare parameters
$\sigma$ and $\lambda$ depends on the way these are renormalized, and
this may be studied close to $d = 2$. It is simpler to work in the
unshifted version (\ref{heffns}), where it is straightforward to
identify the most singular (`bubble') diagrams in powers of
$\epsilon^{-1}$ at a given order in $\lambda$. The mass in the shifted
DP hamiltonian (\ref{heffsh}) then becomes 
$\mu_R - \sigma_R = \sigma (I_d-1) / (I_d+1)$, where
$I_d \equiv (C_d \lambda / D \epsilon) 
[(\mu_R + \sigma_R) / D]^{-\epsilon/2} = (C_d \lambda / D \epsilon)
(\sigma / D)^{-\epsilon / 2}$ because $\mu_R + \sigma_R = \sigma$ in
this approximation. For sufficiently small $\sigma$, this is positive,
indicating that the system is in the inactive phase with an 
{\it exponential} decay of the density. The transition to the active
phase occurs at 
$\sigma_c = D (C_d \lambda / D \epsilon)^{2 / \epsilon}$. Although
this result is accurate only to leading order in $\epsilon$, the
general feature of a transition at a finite value of $\sigma$ in the
DP universality class should persist to $d = 1$. For $d = 2$ the 
transition is seen to continue to occur at a finite value 
$\sigma_c \sim D e^{-4 \pi D / \lambda}$, as $\lambda \to 0$. However,
for larger $d$, the annihilation rate $\lambda$ which drives the
generation of the process $A \to \oslash$, essential for the DP
inactive state, becomes irrelevant, and one may use the same set of
diagrams to argue that there is now no transition, at least for small
$\lambda$.

The same result can be obtained in the framework of an RG calculation
similar to that invoked for $m$ even. This method yields for $m = 3$
the same qualitative picture as for $m = 1$, but the transition moves
closer to the mean field critical point, with 
$\sigma_c \sim D e^{- 5.68 \pi D / \lambda}$ \cite{cartau}, and we
expect this tendency to hold for larger odd values of $m$, in accord
with numerical simulations \cite{taktre,jensen}.

To summarize, we have provided the first analytic theory of branching
and annihilating random walks which explains most of their observed 
behavior. We have shown that the fluctuations responsible for the
failure of mean field theory in the pure annihilation process for
$d \leq 2$ are also responsible for shifting the critical value of the 
branching rate away from zero. For $m$ odd this occurs for all 
$d \leq 2$, with the subsequent transition being in the DP
universality class, while for $m$ even this effect is postponed to
lower dimensions $d < d_c'\approx 4/3$. Our theory correctly takes
account of the symmetry in this case, but is so far unable to yield
accurate estimates of the critical exponents in $d = 1$. However, a
truncated loop expansion appears to provide at least a qualitative
picture of the transition. It would of course be desirable to find
some other controlled approximation scheme in which to approach this
problem. Our investigation of an $N$ species generalization of the
$m=2$ reaction failed to provide us with additional insight in the
single species case, but instead uncovered yet another new
universality class for $N > 1$, with $\sigma_c = 0$ and governed by
the exponents of the exactly solvable $N \to \infty$ limit. This
underlines the importance of fluctuations and correlation effects in
reaction--diffusion systems at low dimensions, which may lead to
remarkably rich nonequilibrium phase diagrams.

We benefitted from discussions with M.~Droz, G.~Grinstein,
N.~Menyh\'ard, K.~Oerding, Z.~R\'acz, and G.M.~Sch\"utz. This research
was supported by the \hbox{EPSRC} through Grant GR/J78327.

\end{multicols}

\end{document}